\PassOptionsToPackage{warn}{textcomp}
\PassOptionsToPackage{table}{xcolor}

\documentclass[a4paper]{article}

\usepackage{amsthm}
\usepackage{amsmath}
\usepackage[affil-it]{authblk}

\usepackage{amsthm}
\usepackage{amsmath}
\usepackage[affil-it]{authblk}

\usepackage[english]{babel}

\usepackage{booktabs} 
\usepackage{paralist}
\usepackage{graphicx}

\usepackage{subfig}
\usepackage{textcomp}
\usepackage{xspace}
\usepackage[table]{xcolor}

\usepackage{multicol}
\usepackage{multirow}
\usepackage{latexsym}
\usepackage{amssymb,amsfonts}
\usepackage{verbatim}
\usepackage{booktabs}
\usepackage[final]{listings}
\usepackage{fullpage}

\usepackage[ruled]{algorithm2e} 
\usepackage[utf8]{inputenc}
\usepackage{url}
\usepackage{hyperref}
\usepackage[numbers,sort]{natbib}

\theoremstyle{definition}
\newtheorem{example}{Example}

\lstdefinestyle{interfaces}{
  float=tp,
  floatplacement=tbp,
  abovecaptionskip=-5pt
}

\begin{document}

\title{The SPECIAL-K Personal Data Processing Transparency and Compliance Platform}

\author[1]{Sabrina Kirrane}
\author[1]{Javier D. Fernández}
\author[2]{Piero Bonatti}
\author[3]{Uros Milosevic}
\author[1]{Axel Polleres}
\author[4]{Rigo Wenning}

\affil[1]{Vienna University of Economics and Business, Austria}
\affil[2]{Universita di Napoli Federico II, Naples, Italy}
\affil[3]{Tenforce, Belgium}
\affil[4]{W3C, Sophia-Antipolis, France}

\date{Dated: \today}

\maketitle

\begin{abstract}
The European General Data Protection Regulation (GDPR) brings new challenges for companies who must ensure they have an appropriate legal basis for processing personal data and must provide transparency with respect to personal data processing and sharing within and between organisations. Additionally, when it comes to  consent as a legal basis, companies need to ensure that they comply with usage constraints specified by data subjects. This paper presents the policy language and supporting ontologies and vocabularies, developed within the SPECIAL EU H2020 project, which can be used to represent data usage policies and data processing and sharing events. We introduce a concrete transparency and compliance architecture, referred to as SPECIAL-K, that can be used to automatically verify that data processing and sharing complies with the data subjects consent. Our evaluation, based on a new compliance benchmark, shows the efficiency and scalability of the system with increasing number of events and users.
\end{abstract}


\section{Introduction}
\label{s:introuction}

The European General Data Protection Regulation (GDPR) defines a set of obligations for controllers and processors of personal data. Primary obligations include ensuring an appropriate legal basis prior to processing personal data and providing transparency to data subjects with respect to processing and sharing of their personal data.
With the coming into effect of the GDPR in May 2018, several tools that can be used to assist companies to assess the compliance of their systems and processes with respect to obligations set forth in the GDPR have been proposed (cf., \cite{ICO2017,MicrosoftTrustCenter2017,Nymity}). 
However, these tools are targeted at self assessment (i.e. companies complete standard questionnaires in the form of a privacy impact assessment) and cannot be used to automatically check compliance of existing systems. 

In particular, when companies rely on consent as a legal basis, such transparency and compliance mechanisms require not only machine-readable representations of the users consent, but also machine-readable representations of data processing and sharing events. SPECIAL\footnote{\url{https://www.specialprivacy.eu/}} is an EU H2020 research and innovation action, which addresses these challenges by demonstrating how semantic technologies can be used to: (i) model data usage policies using the SPECIAL usage policy language; and (ii) represent data processing and sharing events in a semantic log. Both of which have been developed in close collaboration with legal experts, thus ensuring that our automated compliance checking is tightly coupled with the legal assessment process. In addition, we propose the SPECIAL-K architecture, a scalable solution that can be used to log personal data usage policies and events in a manner that supports automated compliance testing. In order to ensure a thorough evaluation of our platform, and to support future comparative analysis, we propose the SPECIAL transparency and compliance benchmark. Our evaluation on synthetic policies and events shows that SPECIAL-K scales with increasing number of users both in a streaming and a batch scenario.
Summarising our contributions:
\begin{itemize}
    \item We demonstrate how the SPECIAL policy language can be used to represent data usage policies. We provide an initial taxonomy for the components of the policy (data categories, processing, etc.) for use in a variety of use cases across multiple domains (e.g., finance, media, insurance), which can be further extended for concrete use cases; 
    \item We define the SPECIAL event log vocabulary, which is derived from the policy language, and show how data processing and sharing events can be used to automatically check compliance with regard to the data subjects consent;
    \item We present the SPECIAL-K Apache Kafka based big data platform, which is able to store logs at scale and to perform scalable compliance checking; and 
    \item We propose a synthetic benchmark for transparency and compliance, which is designed on the basis of well-identified choke points (challenges) that could affect the performance of SPECIAL-K and similar systems.
\end{itemize}

In order to ensure reproducibility and comparative analysis all code that has been developed within SPECIAL is available via a project-specific GitHub space\footnote{\url{https://github.com/specialprivacy}} under an Apache Version 2.0 License\footnote{\url{http://www.apache.org/licenses/}}. Additionally, all SPECIAL ontologies and vocabularies can be accessed via the SPECIAL resources page\footnote{ \url{https://www.specialprivacy.eu/platform/ontologies-and-vocabularies}}.

The remainder of the paper is structured as follows: Section \ref{s:stateArt} discusses alternative policy languages, logging mechanisms and vocabularies, together with GDPR compliance tools.  We present the SPECIAL policy language and the log vocabulary, and introduce the main components of the SPECIAL-K architecture for GDPR transparency and compliance  in Section \ref{s:policy}. We define the SPECIAL compliance benchmark in Section \ref{s:benchmark} and subsequently present the results of the performance evaluation of SPECIAL-K in Section \ref{s:evaluation}. Finally, Section \ref{s:conclusions} concludes and highlights potential avenues for future work. 


\section{Policy Language and Vocabularies}
\label{s:policy}

We first present a motivating use case scenario that exemplifies the requirements derived from the SPECIAL pilots. Following on from this we describe the SPECIAL consent, transparency, and compliance framework. Next, we introduce the SPECIAL policy language and log vocabulary. Finally, we provide a high-level overview of the SPECIAL-K platform.  In the examples that follow, the \texttt{spl} prefix represents \texttt{\small \url{http://www.specialprivacy.eu/langs/usage-policy\#}}. 

\subsection{Motivating Use Case Scenario}

Sue buys a wearable device for fitness tracking from a company called \emph{BeFit}. During the set up, Sue is presented with a \emph{consent request} associated to a data usage policy. The policy states that, in order to provide the tracking service, the device will record biomedical and location data, i.e. the heart rate, duration and location of the fitness activities, and these data will be stored in BeFit's servers in EU. BeFit additionally asks if these data can be used to create an activity profile that can be shared with other companies for marketing purposes (e.g. ads related to fitness including some discount for BeFit users). Sue accepts this option and starts using the device. The signed usage policy is stored in a \emph{transparency log}, together with all processing and sharing events generated from the use of the device by Sue. Two years later, Sue is not using the device anymore and she starts receiving emails from a local gym that advertises its activities. Sue can connect to the log and discover that: (i) BeFit built her profile by mining the data collected by the device; (ii) the profile, stating that she was not doing exercise lately, was shared to the local gym; and (iii) all this was compliant with Sue's data usage policy. At this point, Sue can now decide to revoke the given consent and ask both BeFit and the gym to delete all of her data. 

\begin{figure}[t!]
\center
  \includegraphics[width=\linewidth]{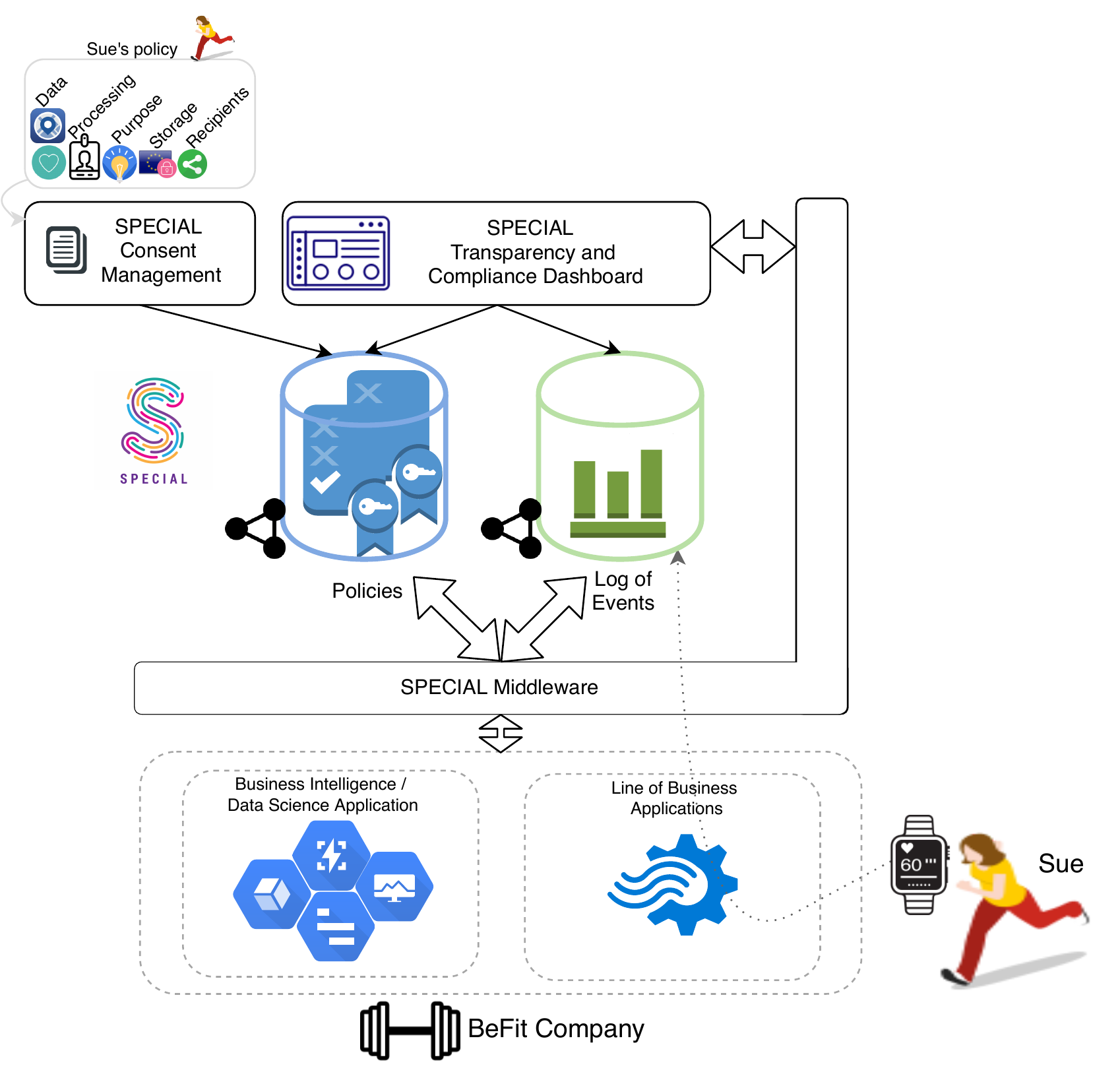}
  \caption{The SPECIAL Consent, Transparency and Compliance Framework.}
  \label{fig:architecture}
\end{figure}

\subsection{Consent, Transparency and Compliance}

The SPECIAL consent, transparency and compliance framework (shown in Figure \ref{fig:architecture}) consists of two primary components: (i) the \emph{SPECIAL Consent Management} component, which is responsible for obtaining consent from the data subject and representing it in the form of a usage policy; and (ii) the \emph{SPECIAL Transparency and Compliance Component}, which is responsible for presenting data processing and sharing events in an easily digestible manner and demonstrating that existing data processing and sharing complies with usage control policies.
\emph{SPECIAL Middleware} includes sub-components that connect the SPECIAL primary components with existing line of business access control mechanisms and business logic and sub-components that enable companies to perform policy aware business intelligence and data science.
In addition to existing data sources that support business operations (i.e. \emph{Line of Business Applications}) and strategic decision making (i.e. \emph{Business Intelligence / Data Science Applications}), we propose two additional data sources, one which is used to store the consent, regulatory and business \emph{Policies} and another to store the data processing or sharing \emph{Events}.

\begin{table*}[t!]
\begin{center}
\caption{SPECIAL auxiliary vocabularies for usage policies.}
\scriptsize
\rowcolors{2}{gray!25}{white}
\begin{tabular}{p{1.3cm} p{3.6cm} p{1.1cm} p{3.8cm} p{2.2cm}}
\rowcolor{gray!50}
\toprule
Category & Namespace & \#Classes & Examples & Superclass\\
\hline
Data & svd:=(S)/vocabs/data & 27 & svd:Activity, svd:Anonymized,\newline svd:Financial, svd:Health,\newline svd:Location, svd:Navigation,\newline svd:Preference, svd:Profile, etc. & spl:AnyData\\
Processing & svpr:=(S)/vocabs/processing & 9 & svpr:Aggregate, svpr:Analyze,\newline svpr:Anonymize, svpr:Collect,\newline svpr:Copy, svpr:Derive, \newline svpr:Move, svpr:Query, \newline svpr:Transfer & spl:AnyProcessing\\
Purpose & svpu:=(S)/vocabs/purposes & 31 & svpu:Account, svpu:Arts,\newline svpu:Delivery, svpu:Education,\newline svpu:Feedback, svpu:Gaming,\newline svpu:Health,svpu:Marketing, svpu:Payment, svpu:Search, etc. &  spl:AnyPurpose\\
Recipient & svr:=(S)/vocabs/recipients & 6 & svr:Delivery, svr:OtherRecipient,\newline svr:Ours, svr:Public, svr:Same, \newline svr:Unrelated & spl:AnyRecipient\\
Storage \newline location & svl:=(S)/vocabs/locations & 7 & svl:ControllerServer, svl:EU,\newline svl:EULike,\newline svl:ThirdCountries, svl:OurServers,\newline svl:ProcessorServers, svl:ThirdParty &  spl:AnyLocation \\
Storage \newline duration & svdu:=(S)/vocabs/duration & 4 & svdu:BusinessPractices, svdu:Indefinitely, svdu:LegalRequirement, svdu:StatedPurpose & spl:AnyDuration \\
\end{tabular}
\label{t:auxiliaryVocabularies}
\normalsize
\end{center}
\end{table*} 

\subsection{The SPECIAL Policy Language}
\label{ss:MCM}

Conceptually, a \emph{usage policy} is meant to specify a \emph{set of
authorised operations}. According to the GDPR, these policies shall specify clearly which data are collected, what is the purpose of the collection, what processing will be performed, and whether or not the data will be shared with others. As such SPECIAL usage policies consist of the following five elements: 

\begin{description}

\item[Data] describes the personal data collected from the data subject. In order to describe which categories of data are collected, an ontology of \emph{personal data} is needed to cover the most common data categories. It is envisaged that the ontology will be extended with suitable profiles and/or integrated with further use case specific ontologies.
\item[Processing] describes the operations that are performed on the personal data. Data processing should be described through a suitable ontology of data operations. 
\item[Purpose] specifies the objective that is associated with data processing. 
Objectives such as marketing, service optimisation and personalisation, scientific
research, are pervasive across a variety of contexts. Purpose descriptions are part of most usage policy languages developed so far (e.g. P3P \cite{DBLP:books/daglib/0007179} and ODRL  \cite{ODRL}). 

\item[Storage] specifies where data are stored and for how long. Note that the GDPR requires that storage is strictly bound to the service needs. This implies storage
minimisation, hence the need to express \emph{upper bounds} to storage
duration, that may be expressed either in terms of the duration of the
service that the data have been collected for, or in absolute terms.
\item[Recipients] specifies who is going to receive the results of
data processing and, as a special case, whom data are shared with. The GDPR does not clearly state to which level of detail this information has to be specified, and there are potentially conflicting needs, such as the companies' desire to keep some of their business relations confidential, and the data subjects' right to trace the flow of their personal information. 
\end{description}

Table \ref{t:auxiliaryVocabularies} provides a high level overview of the initial vocabularies that are necessary to represent the elements of the SPECIAL minimum, core usage policy model. All namespaces share the \texttt{S} which represents \texttt{\small http://www.specialprivacy.eu/}. Note that these vocabularies have been developed to support the initial SPECIAL use cases. Further terms will be added to accommodate additional use cases as needed.
For this purpose, SPECIAL setup the W3C Data Privacy Vocabularies and Controls Community Group (DPVCG) in 2018. The group launched on the 25th of May 2018, the official start date of the GDPR. The mission of the DPVCG is to develop a taxonomy of privacy terms, which include in particular terms from the new European General Data Protection Regulation (GDPR), such as a taxonomy of personal data as well as a classification of purposes (i.e., purposes for data collection), and events of disclosures, consent, and processing such personal data.
In 2019, the group published their first versions of the Data Privacy Vocabulary\footnote{https://www.w3.org/ns/dpv} and the DPVCG GDPR Legal Basis Vocabulary\footnote{https://www.w3.org/ns/dpv-gdpr}. Additional details can be found in \cite{pandit2019creating}.

\subsubsection{Basic Usage Policies}

A usage policy is composed of one or more \emph{basic usage policies},
each of which is an OWL~2 expression of the form\footnote{For the policy language examples we use the functional syntax which is less verbose.}:
\begin{equation}
\scriptsize
  \label{bup}
  \begin{minipage}[t]{0.9\textwidth}
    \tt
    ObjectIntersectionOf(
    
    ~~~ObjectSomeValuesFrom(\textbf{spl:hasData} {\it SomeDataCategory})

    ~~~ObjectSomeValuesFrom(\textbf{spl:hasProcessing} {\it SomeProcessing})
    
    ~~~ObjectSomeValuesFrom(\textbf{spl:hasPurpose} {\it SomePurpose})
    
    ~~~ObjectSomeValuesFrom(\textbf{spl:hasRecipient} {\it SomeRecipient})
    
    ~~~ObjectSomeValuesFrom(\textbf{spl:hasStorage} {\it SomeStorage})
    
    )
  \end{minipage}
\end{equation}
 
The important parts in this expression are the policy's attributes highlighted in bold. The policy author needs to decide for each of them a suitable range highlighted in italics. The example presented authorizes all operations that:
\begin{inparaenum} [(i)]
\item fall within the specified \textit{SomeProcessing} category,
\item operate only on data that belong to \textit{SomeDataCategory},
\item have any purpose covered by the \textit{SomePurpose} category,
\item disclose the results to any member(s) of the \textit{SomeRecipient} category, and
\item store the results in any place belonging to the \textit{SomeStorage} category.
\end{inparaenum}

Therefore, policy (\ref{bup}) encodes the set of all authorisations that have (at least) the specified attributes, which match the minimum core model introduced in the previous section.
Although SPECIAL defines auxiliary vocabularies providing a set of classes for \textit{SomeDataCategory, SomeProcessing, SomePurpose, SomeRecipient}, it should be noted that it is not possible to enumerate over all possible classes and as such the policy language and by extension the vocabularies were designed to be extensible.

 \subsubsection{General Usage Policies}
 
A general usage policy may contain a union of any number of basic policies, each of them of the form (\ref{bup}).  
The resulting policy is conceptually the \emph{union} of all the
authorisations supported by the basic policies, that is, an operation
is authorised by the general policy if and only if the operation is
authorized by \emph{at least one} of its basic policies. 
For instance, the following \emph{general usage policy} states that personal data can only be used for non-commercial purposes and shall neither be stored nor disclosed to third parties, while pseudonymised data can be used freely (where auxiliary vocabularies define the terms \textit{PersonalData, NonCommercial, PseudonymizedData}):

\begin{equation}
\scriptsize
  \label{gup}
  \begin{minipage}[t]{.9\textwidth}\scriptsize
    \tt
    ObjectUnionOf(
    
    ~~ObjectIntersectionOf(
    
    ~~~~ObjectSomeValuesFrom(\textbf{spl:hasData} {\it PersonalData})

    ~~~~ObjectSomeValuesFrom(\textbf{spl:hasProcessing} {\tt spl:AnyProcessing})
    
    ~~~~ObjectSomeValuesFrom(\textbf{spl:hasPurpose} {\it NonCommercial})
    
    ~~~~ObjectSomeValuesFrom(\textbf{spl:hasRecipient} {\tt spl:Null})
    
    ~~~~ObjectSomeValuesFrom(\textbf{spl:hasStorage} {\tt spl:Null})
    
    ~~)

    ~~ObjectIntersectionOf(
    
    ~~~~ObjectSomeValuesFrom(\textbf{spl:hasData} {\it PseudonymizedData})

    ~~~~ObjectSomeValuesFrom(\textbf{spl:hasProcessing} {\tt spl:AnyProcessing})
    
    ~~~~ObjectSomeValuesFrom(\textbf{spl:hasPurpose} {\tt spl:AnyPurpose})
    
    ~~~~ObjectSomeValuesFrom(\textbf{spl:hasRecipient} {\tt spl:AnyRecipient})
    
    ~~~~ObjectSomeValuesFrom(\textbf{spl:hasStorage} {\tt spl:AnyStorage})
    
    ~~)

    )
  \end{minipage}
\end{equation}

\subsubsection{Use Case Specific Usage Policies}

Considering our motivating usecase scenario, Example \ref{ex:sue} demonstrates what Sue's policy would look like if it were represented in the SPECIAL policy language. In this example, the auxiliary vocabularies need to be extended with three new classes: the class \texttt{\small ex:HeartRate} (as a subclass of \texttt{\small svd:Health}), \texttt{\small ex:Profiling} (a subclass of \texttt{\small svpr:Analyze}) and  \texttt{\small ex:Recommendation} (a subclass of \texttt{\small svpu:Marketing}).

\begin{example}
\label{ex:sue}
The following policy: \emph{``Heart rate  and location data are collected and analysed to create a user profile for the purpose of issuing recommendations. Profiles are stored indefinitely in the EU by the data controller and released to third parties."} can be formalised as follows with a factorised general policy:
\begin{center}\scriptsize
  \renewcommand{\t}{\hspace*{1em}}
  \newcommand{\blue}[1]{\textcolor{blue}{#1}}
  \begin{minipage}{.95\textwidth}
    \tt
    ObjectIntersectionOf(\\
    \t ObjectSomeValueFrom( \textbf{spl:hasData} \\
    \t \t \blue{ObjectUnionOf(} \\
    \blue{ 
      \t \t \t ex:HeartRate svd:Location
     )})\\
    \t ObjectSomeValueFrom( \textbf{spl:hasProcessing}
    \blue{ex:Profiling} )\\ 
    \t ObjectSomeValueFrom( \textbf{spl:hasPurpose}
    \blue{ex:Recommendation} )\\ 
    \t ObjectSomeValueFrom( \textbf{spl:hasStorage} \\
    \t \t ObjectIntersectionOf( \\
    \t \t \t ObjectSomeValuesFrom( \textbf{spl:hasLocation} \\
    \t \t \t \t \blue{ObjectIntersectionOf( svl:OurServers svl:EU )})\\
    \t \t \t DataSomeValuesFrom(  ~~\textbf{spl:durationInDays} \\
    \t \t \t \t DatatypeRestriction( xsd:integer \\
    \t \t \t \t \t \textbf{xsd:mininclusive} \blue{"0"}\verb|^^|xsd:integer ))))\\ 
    \t ObjectSomeValueFrom( \textbf{spl:hasRecipient}
    \blue{spl:AnyRecipient} )\\ 
    )
  \end{minipage}
\end{center}
\qed
\noindent
\end{example}

\subsection{The SPECIAL Log Vocabulary}

For the SPECIAL log vocabulary we followed the large body of work in the Business Process Management (BPM) community that focuses on using process execution events for business process compliance monitoring \cite{ly2015compliance}.
The SPECIAL log vocabulary, depicted in Figure \ref{fig:outline}, is composed of the following core aspects: (i) a log entry contains data related to a \emph{single process} and events are instantaneous thus they are associated with a specific timestamp; (ii) in order to represent BPM information (i.e. cases, processes and activities) we integrated an optional \emph{BPM} module that might be present in the company and can complement the logging information;  and (iii) we also integrated an optional \emph{Immutable} module to demonstrate that a log entry can additionally be linked to an immutable record of the event, potentially stored in a different ledger or knowledge base. 

\subsubsection{Logs and Log Entries}

A log (represented as {\tt\small splog:Log}) is a collection of data that records data processing and sharing events as well as consent-related activities (assertion and revocation). The log contains: (i) {\em general log metadata} that describe the log as a whole, such as the data processor whose service is logged, modelled with the {\tt\small splog:processor} property (a subproperty of {\tt\small prov:agent}); and (ii) {\em log entries} ({\tt\small splog:LogEntry}), linked via the {\tt\small splog:logEntry} property (a {\tt\small prov:wasGeneratedBy} subproperty).

\begin{figure*}[t!]
\center
  \includegraphics[width=0.8\linewidth]{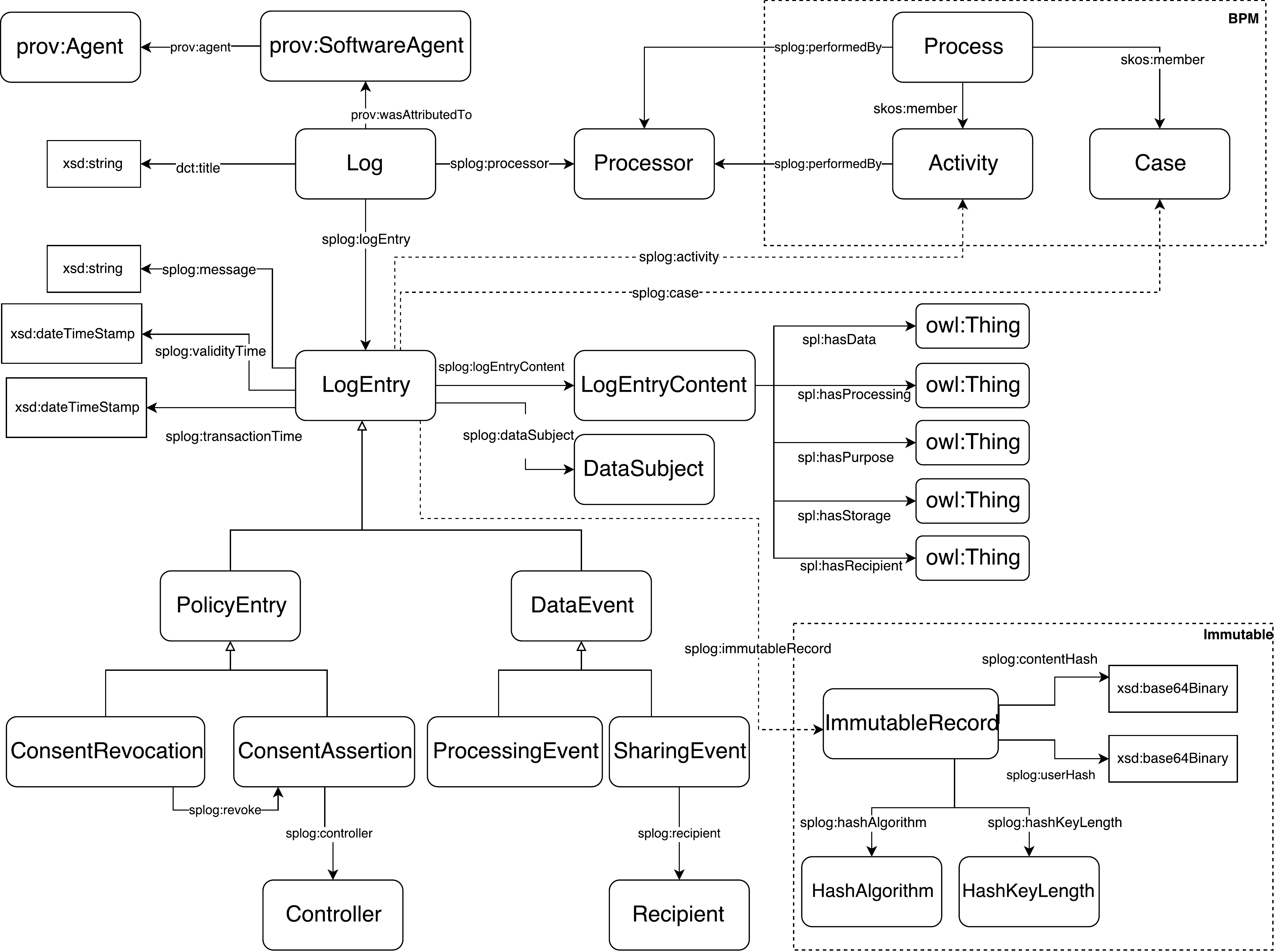}
  \caption{Outline of the SPLog main terms and their relationships}
  \label{fig:outline}
\end{figure*}

Log entries contain information about processing and sharing events associated with data subjects, as well as actions related to the consent provided (or revoked) by the data subject. These different types of entries are represented with a hierarchy of classes, shown in Figure \ref{fig:outline}. Thus, a {\tt\small splog:LogEntry} has two main types (subclasses), {\tt\small splog:PolicyEntry} and {\tt\small splog:DataEvent}, described as follows: 

\begin{description}
\item[PolicyEntry:] This class reflects log entries related to policies and consent. We currently consider two subclasses, {\tt\small splog:ConsentAssertion} specifying a consent provided by a data subject to a {\tt\small splog:Controller} (which in turn can be reachable via the {\tt\small splog:controller} property), and {\tt\small splog:ConsentRevocation}, denoting the revocation of a given consent. Note that we assume that a consent provided by a data subject replaces any previous consent, which can be optionally linked via the {\tt\small splog:revoke} property in our model. 
\item[DataEvent:] This class considers log entries that are actually events on the data, i.e., the aforementioned data processing and sharing events. In the case of the latter, the concrete {\tt\small splog:Recipient} instances can be specified, via  {\tt\small splog:recipient}.
\end{description}

Besides general metadata and a human-friendly message ({\tt\small splog:message}),  the data in a log entry can be described using the following:

\begin{description}
\item[Data subjects:] The log entry \emph{SHOULD} reference the data subject(s) involved in the entry using the {\tt\small splog:dataSubject} property (a {\tt\small prov:wasAssociatedWith} subproperty). Note that in case of anonymised logs, no subject can be specified.
\item[Content:] The log entry \emph{MUST} reference the actual data of the log. This is specified with the {\tt\small splog:logEntryContent} property, which points to the appropriate instance of {\tt\small splog:LogEntryContent}, described below.
\item[Timestamps:] The log entry \emph{MUST} reference the time at which the event occurred using the {\tt\small splog:validityTime} property (subproperty of {\tt\small prov:atTime}). The log entry \emph{SHOULD} also reflect the time in which the log was recorded, using {\tt\small splog:transactionTime} (a {\tt\small dct:issued} subproperty).
\end{description}

Optionally, the entry \emph{MAY} reference a {\tt\small splog:InmutableRecord} of its contents and the concrete BPM {\tt\small splog:Activity} and {\tt\small splog:Case} involved in the process.

\subsubsection{Use Case Specific Log Entry Content}
\label{ss:EventContent}

The {\tt\small splog:LogEntryContent} class describes the actual data usage using the minimum core model (i.e. data, processing, purpose, storage and recipients). Thus event content and data policy authorisations are described with the same class formalisation, which facilitates compliance checking. As such, the {\tt\small splog:LogEntryContent} class definition  \emph{MUST} include the minimum core model elements using the properties {\tt\small spl:hasData}, {\tt\small spl:hasProcessing}, {\tt\small spl:hasPurpose}, {\tt\small spl:hasStorage}, {\tt\small spl:hasRecipient}  defined in the SPECIAL policy language. 

The following example provides an overview of how the SPECIAL Policy Log vocabulary might be used to represent a log. We make use of our BeFit scenario: we assume (i) Sue is using a wearable appliance for fitness tracking from BeFit; (ii) the application is tracking the location of Sue for \emph{health} purposes; and (iii) a new location is stored in a particular database (called \emph{BeFitDatabaseEurope}) and reflected in the log (called \emph{BeFitLog}). Let us also assume that the namespace for the BeFit company is {\tt\small befit:} (pointing to the appropriate IRI), being {\tt\small befit:Us} the main reference of the company. We first show the general log description in Listing \ref{exLog1}.
Then, we include a new entry in the log, which is a processing event (uniquely identified as {\tt\small befit:entry3918}) referencing a new tracking position of Sue, shown in Listing \ref{exEvent1}. We assume Sue's unique identifier is {\tt\small befit:Sue}. The collection of the new position took place on the 3rd of January, 2018, at 13:20 (i.e. validity time) and the event was recorded few seconds later (i.e. transaction time).
Note that the log entry {\tt\small befit:entry3918} is an instance of a {\tt\small ProcessingEvent}, {\tt\small befit:iRec3918} links the immutable version of the event, and {\tt\small befit:content3918} points to the actual content of the event, defined in the following Listing \ref{exContent1}.

\begin{lstlisting}[float=*, frame = single,basicstyle=\scriptsize,caption=Log description for BeFit devices.,label=exLog1]
befit:BeFitLog a splog:Log;
  dct:title "Log of BeFitDatabaseEurope"@en;
  dct:description "This contains events on BeFitDatabaseEurope  tracking devices geo-located 
  		in Europe"@en;
  dct:issued "2018-02-14"^^xsd:date;
  prov:wasAttributedTo befit:BeFitDatabaseEurope ;
  splog:processor befit:Us .    
\end{lstlisting}

\begin{lstlisting}[float=*, frame = single,basicstyle=\scriptsize,caption=A new event for Sue's BeFit device.,label=exEvent1]
befit:BeFitLog splog:logEntry befit:entry3918 . 
        
befit:entry3918 a splog:ProcessingEvent;
 dct:title "Collection of new device position"@en;
 splog:dataSubject befit:Sue ;
 dct:description  "We collected a new position of your BeFit device in our database 
 		in Europe"@en;
 splog:transactionTime "2018-01-10T13:20:50Z"^^xsd:dateTimeStamp;
 splog:validityTime "2018-01-10T13:20:00Z"^^xsd:dateTimeStamp;
 splog:message "Tracking position by GPS... collected!" ;
 splog:eventContent befit:content3918 ;
 splog:inmutableRecord befit:iRec3918 .    
\end{lstlisting}

\begin{lstlisting}[float=*, frame = single,basicstyle=\scriptsize,caption=The content of a new event for Sue's BeFit device.,label=exContent1]
befit:content3918 a splog:LogEntryContent;
dct:description "Location data are collected by a BeFit device only for the health purpose 
		of the service"@en;
 spl:hasData svd:Location;
 spl:hasProcessing befit:SensorGathering;
 spl:hasPurpose befit:HealthTracking;
 spl:hasStorage [has:location svl:OurServers];
 spl:hasRecipient [a svr:Ours].
    
befit:SensorGathering rdfs:subClassOf svpr:Collect .
befit:HealthTracking rdfs:subClassOf svpu:Health .
\end{lstlisting}

\begin{figure}[t!]
\begin{center}
  \includegraphics[width=\linewidth]{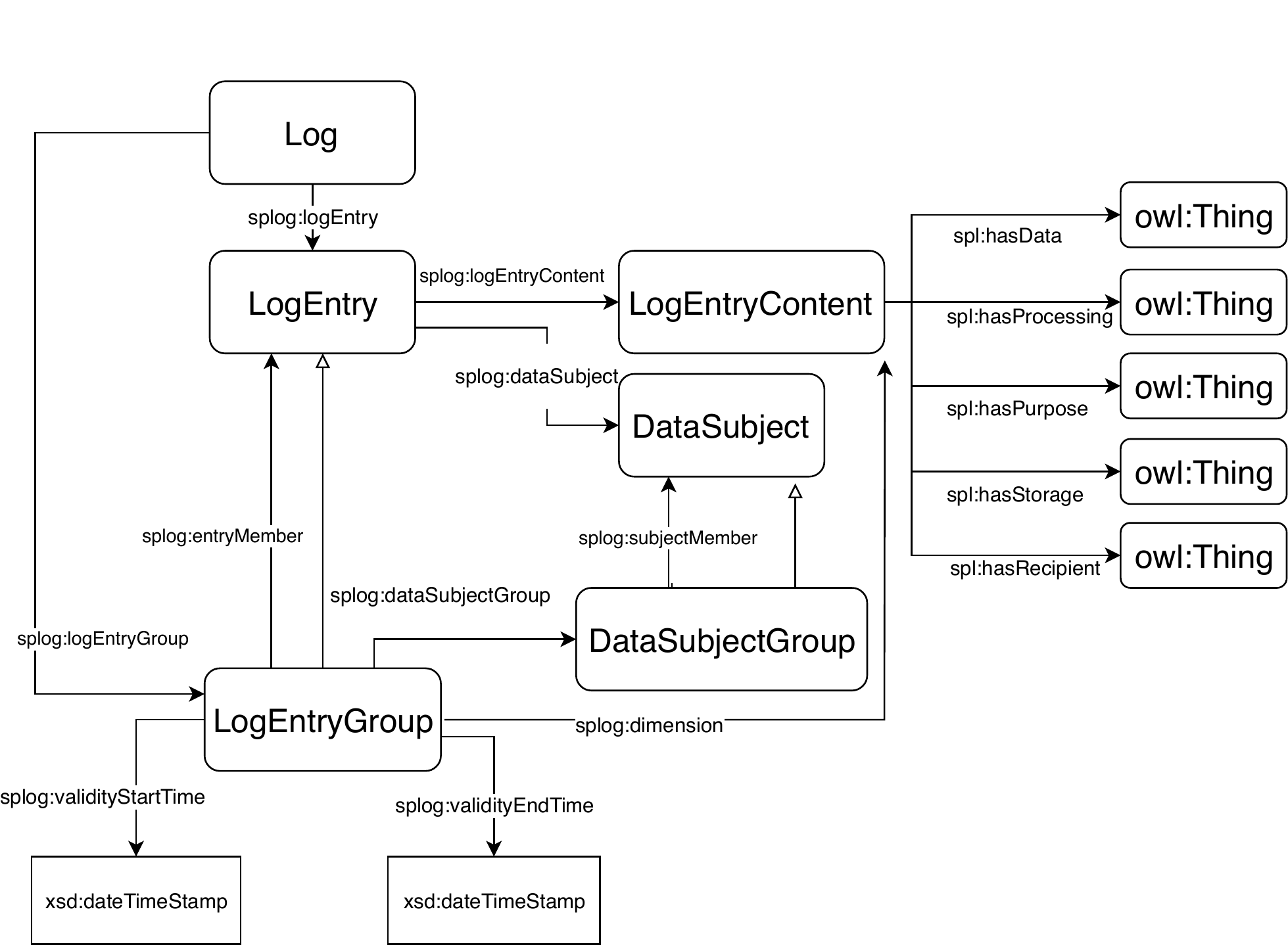}
  \caption{Pictorial summary of log entry grouping}
  \label{fig:outlineGrouping}
\end{center}
\end{figure}

\subsubsection{Grouping Log Entries}

Log entries can be grouped to facilitate scalability in those scenarios where there exists a continuous flow of information, such as the envisioned big data applications. For instance, in our BeFit use case, a log group could be used to represent (as a single entry) the collection of data during a running activity of a data subject in BeFit. 
SPECIAL provides such a grouping model, outlined in Figure \ref{fig:outlineGrouping}. The core class is {\tt\small splog:LogEntryGroup} (a subclass of {\tt\small splog:LogEntry}), which has a validity time interval denoted by the {\tt\small splog:validityStartTime} and {\tt\small splog:validityEndTime} properties (subproperties of {\tt\small prov:startedAtTime} and {\tt\small prov:endedAtTime}, respectively). The group \emph{MUST} reference the content (data, purpose, processing, etc.) it groups via the {\tt\small splog:dimension} property (a {\tt\small splog:logEntryContent} subproperty), which points to a particular {\tt\small splog:LogEntryContent}.  The group \emph{MAY} reference the data subject(s) in the group (all sharing the same log entry content), using the property {\tt\small splog:dataSubjectGroup} ({\tt\small prov:wasAssociatedWith} subproperty). This property points to a {\tt\small splog:DataSubjectGroup} instance that groups all the data subject members in the group via {\tt\small splog:subjectMember} (a {\tt\small skos:member} subproperty). Finally, the group \emph{MAY} point to the particular entries included in the group through the {\tt\small splog:entryMember} property (a {\tt\small skos:member} subproperty). This option can facilitate a fine-grained traceability at the cost of storing additional information (i.e. all log entries of the group), hence it is an optional feature.
Listing \ref{exGrouping1} shows a log grouping the category of recommendations given to Sue, John and Rose during a month.

\begin{lstlisting}[float=*, frame = single,basicstyle=\scriptsize,caption=A grouping example merging all recommendations given in a month.,label=exGrouping1]
befit:BeFitLog a splog:Log ;
 splog:logEntryGroup befit:recommendationsJanuary2018 .
    
befit:recommendationsJanuary2018 a splog:logEntryGroup
 splog:transactionTime "2018-02-01T00:05:00Z"^^xsd:dateTimeStamp;
 splog:validityTime "2018-01-31T23:59:59Z"^^xsd:dateTimeStamp;
 splog:dataSubjectGroup  befit:basicSubjectGroup;
 splog:dimension befit:templateOfferRecommendation .

befit:basicSubjectGroup splog:member befit:Sue,befit:John,befit:Rose.
            
befit:templateOfferRecommendation a splog:LogEntryContent ;
 spl:hasData befit:OfferRecommendation;
 spl:hasProcessing befit:MonthlyDataAnalysis;
 spl:hasPurpose befit:MonthlyOffersRecommendation;
 spl:hasStorage [has:location svl:OurServers];
 spl:hasRecipient [a svr:Ours].

befit:OfferRecommendation rdfs:subClassOf svd:Location;
  rdfs:comment  "We recommended you an offer at the end of the month based on the location 
    of your device" .
    
befit:MonthlyDataAnalysis rdfs:subClassOf svpr:Analyze .
befit:MonthlyOffersRecommendation rdfs:subClassOf befit:RecommendationActivity .
befit:RecommendationActivity rdfs:subClassOf svpu:Marketing .
       
\end{lstlisting}

\subsection{Using SPECIAL Resources for Compliance Checking}

Policies and log events are described in semantically unambiguous terms aligned to the same taxonomies
that are used to define usage policies, hence facilitating transparency and automatic compliance checking. Regarding the latter, the usage policy enforced by a data controller contains the operations that are permitted within the data controller's organisation. Therefore,  the usage $U_c$ attached to a SPECIAL log entry \emph{complies} with the usage policy $P_s$ in the data subject's
consent if and only if all the authorisations in $U_c$
are also authorised by $P_s$, that is, $U_c$ complies with $P_s$ if and only if
  $U_c\subseteq P_s $.
%
Thus, in OWL~2 terminology, this amounts to checking whether the following
axiom is \emph{entailed} (implied) by the combined ontology $\O$ containing the SPECIAL policy language
ontology plus the aforementioned auxiliary vocabularies:
  \mbox{\tt\small SubClassOf(\textit{U}$_c$ \textit{P}$_s$).} This is inherently supported by general inference engines for OWL 2 (e.g. HermiT and FaCT++).
%
For instance, considering a log entry relating to {\tt\small befit:SensorGathering}, such as that presented in \emph{Listing} \ref{exContent1}. This entry is compliant with a potential usage policy stating that the controller can collect ({\tt\small svpr:Collect}) location data \emph{iff}  {\tt\small befit:SensorGathering} is a sublass of {\tt\small svpr:Collect}. 



\subsection{The SPECIAL-K Architecture}

\begin{figure*}[t!]
\center
    \includegraphics[width=0.8\linewidth]{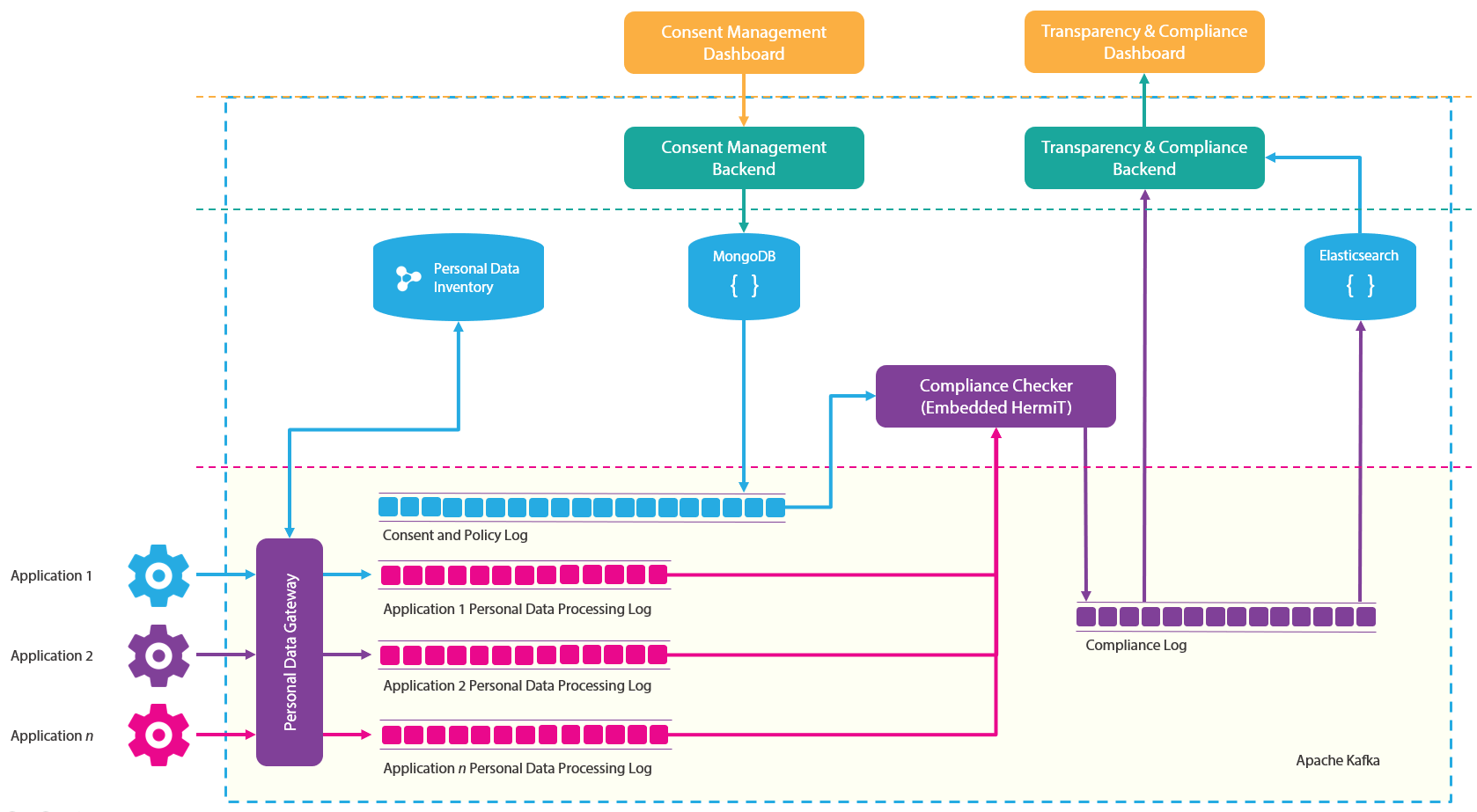}
   \caption{The SPECIAL-K architecture.}
   \label{fig:architecture-post-overview}
\end{figure*}

The SPECIAL platform depicted in Figure \ref{fig:architecture-post-overview} consists of the following components: 
\begin{inparaenum}[(i)]
\item \emph{The SPECIAL Consent Management Dashboard and Backend} is responsible for obtaining consent from the data subject and representing it using the SPECIAL usage policy vocabulary;
\item \emph{The SPECIAL Transparency \& Compliance Dashboard and Backend} is responsible for presenting data processing and sharing events to the user in an easily digestible manner following the SPECIAL policy log vocabulary; and demonstrating that data processing and sharing complies with usage control policies; and
\item \emph{The SPECIAL Personal Data Gateway} enables applications to submit their personal data processing requests, which are subsequently inspected for compliance by the  \emph{Compliance Checker}. 
\end{inparaenum}
 
Data processing and sharing event logs are stored in the Kafka\footnote{https://kafka.apache.org/} distributed streaming platform. A Kafka topic is used to store application logs, while a separate compliance topic (called \emph{Compliance Log}) is used to store the enriched log after compliance checks have been completed. As logs can be serialised using JSON-LD, it is possible to benefit from the faceting browsing  capabilities of Elasticsearch\footnote{https://www.elastic.co/products/elasticsearch}, and the out of the box visualisation capabilities provided by Kibana.
The compliance checker, which currently includes an embedded HermiT\footnote{http://www.hermit-reasoner.com/} reasoner uses the consent saved in MongoDB, together with the application logs provided by Kafka to check that data processing and sharing complies with the relevant usage control policies. The results of this check are saved onto a new Kafka topic.

\section{The SPECIAL Compliance Benchmark}
\label{s:benchmark}

In this section, we present the choke points used to identify technical difficulties that a compliance benchmark should consider in order to challenge the system under test. Following on from this, we introduce the SPECIAL compliance benchmark tasks.

\begin{table*}[t!]
\caption{SPECIAL Compliance Benchmark Tasks.}
\scriptsize
\begin{tabular}{lllllp{2cm}lll}
\toprule
{\bf Task} & {\bf Subtask} & {\bf Scenario} & {\bf \#Users} & {\bf Event Rate} & {\bf Policies} & {\bf Test Time} & {\bf Pass Ratio} & {\bf Choke Point}\\
\hline
\multirow{5}{*}{C-T1} & C-T1-1 & \multirow{5}{*}{Streaming} & \multirow{5}{*}{1000} & \multirow{5}{*}{1 ev./10s} & 1 policy & \multirow{5}{*}{20 minutes} & \multirow{5}{*}{Random} & \multirow{5}{*}{CCP1,CCP5}\\ 
& C-T1-2 & & & & UNION of 5 p.& & & \\
& C-T1-3 & & & & UNION of 10 p.& & & \\
& C-T1-4 & & & & UNION of 20 p.& & & \\
& C-T1-5 & & & & UNION of 30 p.& & & \\
\hline
\multirow{5}{*}{C-T2}& C-T2-1  & \multirow{5}{*}{Streaming} & 100 & \multirow{5}{*}{1 ev./10s} & \multirow{5}{*}{UNION of 5 p.} & \multirow{5}{*}{20 minutes} & \multirow{5}{*}{Random} & \multirow{5}{*}{CCP2,CCP5}\\ 
& C-T2-2 & &1K & & & & & \\
& C-T2-3 & &10K & & & & & \\
& C-T2-4 & &100K & & & & & \\
& C-T2-5 & &1M & & & & & \\
\hline
\multirow{5}{*}{C-T3} & C-T3-1 & \multirow{5}{*}{Streaming} & \multirow{5}{*}{1000} & \multirow{5}{*}{1 ev./10s} & \multirow{5}{*}{UNION of 5 p.} & \multirow{5}{*}{20 minutes} & 0\% & \multirow{5}{*}{CCP3,CCP5}\\
& C-T3-2 & & & & & & 25\% & \\
& C-T3-3 & & & & & & 50\% & \\
& C-T3-4 & & & & & & 75\% & \\
& C-T3-5 & & & & & & 100\% & \\
\hline

\multirow{5}{*}{C-T4} & C-T4-1 & \multirow{5}{*}{Streaming}&  \multirow{5}{*}{1000} & 1 ev./60s & \multirow{5}{*}{UNION of 5 p.} & \multirow{5}{*}{20 minutes}& \multirow{5}{*}{Random} & \multirow{5}{*}{CCP4,CCP5}\\
& C-T4-2 & & & 1 ev./30s& &  & & \\
& C-T4-3 & & & 1 ev./10s& &  & & \\
& C-T4-4 & & & 1 ev./s& &  & & \\
& C-T4-5 & & & 10 ev./s& &  & & \\
\hline
\multirow{5}{*}{C-T5} &  C-T5-1 & \multirow{5}{*}{Batch}& 100 & \multirow{5}{*}{-} & \multirow{5}{*}{UNION of 5 p.} & 100K events & \multirow{5}{*}{Random} & \multirow{5}{*}{CCP6}\\
& C-T5-2 & &1K & & &1M events& & \\
& C-T5-3 & &10K & & &10M events& & \\
& C-T5-4 & &100K & & &100M events& & \\
& C-T5-5 & &1M & & & 1B events& & \\
\hline

\end{tabular}
\label{t:compTasks}
\vspace{1cm}
\normalsize
\end{table*}

\subsection{Choke Point-based Benchmark Design}
\label{s:chocke}

We design the SPECIAL compliance benchmark following the same methodology as most of the benchmarks developed by the H2020 HOBBIT project \cite{ngomo2016hobbit}. Thus, the development of the benchmark is driven by so-called ``choke-points'', a notion introduced by the Linked Data Benchmark Council (LDBC) \cite{erling2015ldbc,angles2014linked}. A choke-point analysis aims to identify important technical challenges to be evaluated in terms of query workload. This methodology depends on the identification of such workload by technical experts in the architecture of the system under test. Thus, we analysed the SPECIAL platform with the technical experts involved in the SPECIAL policy vocabulary, the transparency and the compliance components. Following this study, we identified the following compliance choke points:

\begin{description}
\item[CCP1 - Different ``complexities'' of policies:] In general, policies can be arbitrarily complex, affecting the overall performance of any compliance checking process. Thus, the benchmark must consider different complexities of policies, reflecting a realistic scenario.

\item[CCP2 - Increasing number of users:] The benchmark should test the ability of the system to efficiently scale and perform as increasing number of users, i.e. data processing and sharing events, are managed. 

\item[CCP3 - Expected passed/fail tests:] In general, the benchmark must consider a realistic scenario where policies are updated, some consents are revoked, and others are updated. The benchmark should provide the means to validate whether the performance of the system depends on the ratio of passed/fail tests in the work load. 

\item[CCP4 - Data generation rates:] The system should cope with consents and data processing and sharing events generated at increasing rates, addressing the ``velocity'' requirements of most big data scenarios.

\item[CCP5 - Performant streaming processing:] The benchmark should be able to test the system in a streaming scenario, where the compliance checking should fulfill the aforementioned requirements of \emph{performance and responsiveness} (i.e., latency).

\item[CCP6 - Performant batch processing:] In addition to streaming, the system must deal with performant compliance checking in batch mode. 

\end{description}

\subsection{Data Generation}
\label{s:datageneration}

In the following we present the SPECIAL compliance benchmark data generator used to test the compliance and transparency performance of the SPECIAL platform.
The data generation considers two related concepts: the usage policies and the data sharing and processing events that are potentially compliant with user consent. 
When it comes to the policies, we distinguish three alternative strategies to generate pseudo random policies:

\begin{itemize}
\item [(a)] Generating policies in the PL fragment of OWL 2, disregarding the SPECIAL minimum core model (MCM);
\item [(b)] Generating random policies that comply to the SPECIAL minimum core model (MCM);
\item [(c)] Generating not fully random (i.e. pilot oriented policies) subsets of the business policies.
\end{itemize}

In this benchmark, we focus on the second alternative, providing a synthetic data generator following the BeFit scenario. 
In addition, the classes in the policies and the log events can come from the standard SPECIAL policy vocabulary, or can be extended with new terms from an ontology. At this stage, we consider the SPECIAL policy vocabulary as the core input.
the SPECIAL compliance benchmark data generator supports the following configuration parameters:

\begin{description}
\item[Generation rate:] The rate at which the generator outputs events. This parameter understands golang duration syntax eg: 1s or 10ms.

\item[Number of events:] The total number of events that will be generated. When this parameters is $<=0$ it will create an infinite stream.

\item[Format:] The serialisation format used to write the events (json or ttl).

\item[Type:] The type of event to be generated: {\em log}, which stands for generating data sharing and processing events, or {\em consent}, which generate new user consents.

\item[Number of policies:] The maximum number of policies to be used in a single consent.

\item[Number of users:] The number of UserID attribute values to generate. 
\end{description}

\subsection{Benchmark Tasks}
\label{s:tasks}

Table \ref{t:compTasks} shows the tasks to be performed by the SPECIAL compliance component in order to cover all choke points identified above. Each task delimits the different parameters involved, such as the scenario (streaming or batch processing), the number of users, etc. 
These parameters follow the choke points, and their values are estimated based on consultation with the SPECIAL pilot partners. Note that all tests set a test time of 20 minutes, which delimits the number of events generated given the number of users and event generation rate in each case.

\section{Evaluation}
\label{s:evaluation}

\begin{table}[t!]
\center
\caption{Characteristic of the cluster.}
\scriptsize
\begin{tabular}{p{2.5cm}p{4cm}}
\toprule
Number of Nodes & 10 \\
CPUs & Each node consists of 4 CPUs per machine (2 cores per CPU)\\
Memory & 16 GB per node \\
Disk Space & 100 GB per node \\
Operating System & CoreOS 2023.5.0 (Rhyolite) \\
Replication Factor & 2. As mentioned this implies that data is written to 2 nodes, enhancing fault-tolerance at the cost of additional space requirements and a minimum time overhead \\
\hline
\end{tabular}
\label{t:cluster}
\normalsize
\end{table} 

\begin{figure}[t]
  \begin{center}
    \includegraphics[width=.7\textwidth]{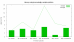}
  \caption{\label{fig:t1-complexPoliciesMedian}Median and average latencies with increasing complex policies.}
    \end{center}
\end{figure}  

\begin{figure}[t]
  \begin{center}
    \includegraphics[width=.7\textwidth]{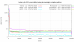}
 \caption{Latencies (in 95\% percentile) with increasing complex policies}
  \label{fig:t1-complexPoliciesLatency}
    \end{center}
\end{figure}  

\begin{figure}[t]
  \begin{center}
    \includegraphics[width=.7\textwidth]{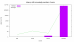}
  \caption{Median and average latencies with increasing number of users.}
  \label{fig:t2-Median}
    \end{center}
\end{figure}  

\begin{figure}[t]
  \begin{center}
    \includegraphics[width=.7\textwidth]{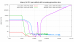}
 \caption{Latencies (in 95\% percentile) with increasing number of users.}
  \label{fig:t2-Latency}
    \end{center}
\end{figure}  

\begin{figure}[t]
  \begin{center}
    \includegraphics[width=.7\textwidth]{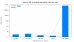}
  \caption[Median and average latencies with increasing generation rates.]{\label{fig:t4-rateMedian}Median and average latencies with increasing generation rates. The rate refers to events per user, with 1K users}
    \end{center}
\end{figure}  

\begin{figure}[t]
  \begin{center}
   \includegraphics[width=.7\textwidth]{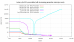}
    \caption[Latencies (in 95\% percentile) with increasing generation rates.]{\label{fig:t4-rateLatency}Latencies (in 95\% percentile) with increasing generation rates. The rate refers to events per user, with 1K users}
    \end{center}
\end{figure}  

\begin{figure}[t]
  \begin{center}
    \includegraphics[width=.7\textwidth]{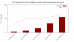} 
  \caption[CPU usage for compliance checking with increasing generation rate.]{CPU usage (in \%) for compliance checking with increasing generation rate (1K users)}
  \label{fig:CPU-usage-t4-rates}
    \end{center}
\end{figure} 

\begin{table*}[t!]
\center
\caption[Space requirements with increasing generation rate]{Space requirements (MB) with increasing generation rate.}
\scriptsize
\begin{tabular}{llrr}
\toprule
{\bf \# Users} & {\bf Event Rate (per user)} & {\bf \# Events} & {\bf Disk Space (MB)} \\
\hline
1,000 & 1 ev./60s & 20,000 &  819 \\
1,000 & 1 ev./30s & 40,000 &  1,563 \\
1,000 & 1 ev./10s & 120,000 &  1,954 \\
1,000 & 1 ev./1s & 1,200,000 &  5,355\\
1,000 & 1 ev./100ms & 12,000,000 &  59,664\\

\hline

\end{tabular}
\label{t:space}
\normalsize
\end{table*} 

\begin{figure}[t]
  \begin{center}
    \includegraphics[width=.7\textwidth]{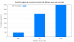}
  \caption[ Total batch compliance checking  throughput with increasing number of compliance checkers.]{\label{fig:batch-throughput-c23-25} Total batch compliance checking  throughput (in events/s) with increasing number of compliance checkers}
    \end{center}
\end{figure}  

\begin{figure}[t]
  \begin{center}
    \includegraphics[width=.7\textwidth]{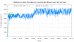}
  \caption[Distribution of batch compliance checking throughput with different users and work load.]{\label{fig:batch-throughput-distribution-c23-25}Distribution of batch compliance checking throughput (in events/s) with different users and work load. We consider 1000 events per user}  
    \end{center}
\end{figure}  

In this section, we use the SPECIAL compliance benchmark to evaluate our SPECIAK-K platform. 
Rather than showing a complete evaluation on an optimised and performant infrastructure, we focus on testing an installation of the SPECIAL platform in order to identify potential optimisations. 
In the following we present the results for all the compliance tasks (\emph{C-T1} to \emph{C-T5} from Table~\ref{t:compTasks}). We disregard \emph{C-T3} as no significant differences were found in our tests and we opt for a more realistic random generation of policies. All experiments were executed on a cluster consisting of 10 nodes. The characteristic of the cluster are presented in \emph{Table}~\ref{t:cluster}. In all cases we report the average results of 3 independent executions.

\subsection{C-T1: Different Complexities of Policies}
\label{ss:t1}

Recall that this task regards the behaviour of the system in a streaming scenario (at 1 event/10s per user and 1K users) when different complexities of policies (measured as the number of union policies) are considered. In this scenario, we make use of 1 compliance checker in order to isolate the performance of one instance. We also compare the implementation of a standard Hermit reasoner with our SPECIAL PLReasoner.
Figure~\ref{fig:t1-complexPoliciesMedian} shows the median and average latencies (in milliseconds) with 1, 10 and 30 union policies. Results show that the median \emph{latency ranges between 1.5-5 ms}, with relatively small differences as the number of union policies grow, except for the union of 30 policies. In this case, the higher number of union policies allows Hermit to quickly find a match (1.5 ms). As for the comparison of reasoners, Hermit seems to slightly outperform our PLReasoner in the scenario under test. Nonetheless, when both reasoners are run in isolation, the tailored PLReasoner engine is several times faster than Hermit \cite{specialD3.5}. A first analysis shows that different parsing and deserialisation of policies can affect the times of PLReasoner in the SPECIAL platform. In addition, our isolated study generally considers richer and more complex policies than the SPECIAL compliance benchmark, also including different time intervals (for the duration of the storage).
Finally, the higher figures for the average latency again denote a skewed distribution. Thus, we inspect the latency at 95\% percentile (the value at which 95\% of the data is included), depicted in Figure~\ref{fig:t1-complexPoliciesLatency} for 1, 10 and 30 policies. The distribution shows that, in all scenarios, the latency at 95\% percentile is stable after the warm-up, with small differences with more union policies. Results also show that only 5\% of the events can experience latencies over 5 ms. 

\subsection{C-T2: Increasing Number of Users}
\label{ss:t2}

The second task in the SPECIAL compliance benchmark focuses on evaluating the scalability of the system with increasing number of users, from 100 to 1 million. These users are considered to be generating events in parallel, each of them at a rate of 1 event every 10 seconds. In the following evaluation, we study the first four subtasks, covering up to 100,000 users given the characteristics of the experimental infrastructure. Note that serving 100,000 users at the aforementioned rate already implies the need to manage a stream of 10,000 events every second. In this scenario, we consider 10 compliance checkers running in parallel in order to cope with such demand. 
%
Figure~\ref{fig:t2-Median} shows the median and average latencies for 100-100,000 users. Results show that the system is able to provide a median latency of less than 1ms with 1,000 users (each user with 1 event every 10 seconds, hence overall the system receives 1 event every 10 ms simultaneously), and 1.6ms with 10,000 users (overall, 1 event very ms). However, with 100,000 users, the current infrastructure needs to manage 1 event every 0.1ms (less than the checking time of 1ms), which causes delays of several seconds. 
In order to highlight a potential worst-case scenarios, we represent the latency at 95\% percentile in Figure~\ref{fig:t2-Latency}. Note that an increasing number of users results in more events, hence the different number of events in each scenario. As expected, results show two different scenarios. On the one hand, a number of users between 100-10,000 results in a \emph{95\% percentile around 1 ms}, with an initial warm-up step that produces higher latencies. On the other hand, a higher number of users (100,000) leads to increasing latencies as the number of events grows, i.e. events are queued for several seconds. The main reason is that the number of compliance checkers (10, given the amount of computational resources in the cluster) cannot cope with the overall actual ratio of 10,000 events every second. Given that Kafka is able to optimise and adapt to the number of parallel checkers, which is solely limited by the number of partitions in the cluster, hence a more powerful infrastructure could cope with a greater number of users.

\subsection{C-T4: Increasing Data Generation Rates}
\label{ss:t4}

This task evaluates the performance of the system with increasing streaming rates. We consider 10 compliance checkers running in parallel in order to try to cope with the biggest rates in the defined tasks.
Figure~\ref{fig:t4-rateMedian} represents the median and average latencies (in milliseconds and logarithm scale), while the latency at 95\% percentile is shown in Figure~\ref{fig:t4-rateLatency} (in logarithm scale). Several comments are in order. First, note that the median values in Figure~\ref{fig:t4-rateMedian} are consistent with our previous latency measures, obtaining values between 1-2ms for rates up to 1 ev/s (per user). Then, as expected, the median latency increases up to several seconds at the highest rate of 1 ev/100ms per user, that is, the system receives a total of 1 ev/0.1ms.
The huge skewed distribution for the highest rate is revealed by the 95\% percentile shown in Figure~\ref{fig:t4-rateLatency}. Note that we fix the benchmark time at 20 minutes, so more events are generated with increasing generation rates. Results show that the latency reaches a stable stage for rates up to 1 ev/1s per user, i.e. a total of 1 ev/1ms. In contrast, the latency at 95\% percentile grows steadily for streams at 1 ev/100ms per user. This fact shows that the current installation cannot cope with such high rates and new events have to queue until they can be processed. The maximum latency reaches 17 minutes for 12 million events. 
Finally, we inspect the CPU usage and the overall disk space of the solution. The CPU usage (in percentage) is represented in Figure~\ref{fig:CPU-usage-t4-rates}. As expected, the results show that the CPU usage increases (but sublinearly) with the generation ratio. The disk space requirements are given in Table~\ref{t:space}. It is worth mentioning that the disk space depends on multiple factors, such as the individual size of the randomly generated events, the aforementioned level of replication, the number of nodes and the level of logging/monitoring in the system. The reported results already show the log compaction feature of Kafka as, on average, less bytes are required to represent each of the events with increasing event rates. 

\subsection{C-T5: Batch Performance}
\label{ss:t5}

Recall that this task considers a batch processing scenario, i.e. events are already loaded in the system, with an increasing number of events and users. In this evaluation, we consider the first three subtasks, testing up to 10 million events (considering 100K events per user). We inspect the provided throughput (processed events per seconds) using an increasing number of compliance checkers. As in previous cases, we here consider 10 compliance checkers running in parallel.
Figure~\ref{fig:batch-throughput-c23-25} shows the total batch throughput (in events/s) for 100K, 1M and 10M events. The total throughput increases with the number of events, being over 474 processed events/s in all cases, with a maximum of 3,489 events/s in the case of 10M events. 
Finally, Figure~\ref{fig:batch-throughput-distribution-c23-25} looks at the distribution of the throughput for the case of 1M and 10M events. Both cases shows similar initial figures, with increased performance around 4M events.

\section{Related Work}
\label{s:stateArt}

When it comes to the representation of usage policies there are several potential candidates including semantic policy languages \cite{DBLP:conf/policy/UszokBJSHBBJKL03,rei,DBLP:journals/tkde/BonattiCOS10,Kolovski2007} and standard based policy languages \cite{DBLP:books/daglib/0007179,ODRL}.
KAoS \cite{DBLP:conf/policy/UszokBJSHBBJKL03} is a general policy language which adopts a pure ontological approach, whereas Rei \cite{rei} and Protune \cite{DBLP:journals/tkde/BonattiCOS10} use ontologies to represent concepts, the relationships between these concepts and the evidence needed to prove their truth, and rules to represent policies. 
Kolovski et al. \cite{Kolovski2007} demonstrate how together description logic and defeasible logic rules can be used to understand the effect and the consequence of sets of access control policies. 
While, the Platform for Privacy Preferences (P3P)\footnote{P3P,\url{http://www.w3.org/TR/P3P/}}, is a W3C recommendation, which enables websites to express their privacy preferences in a machine readable format. A more recent W3C recommendation known as the Open Digital Rights Language (ODRL)\footnote{ODRL,\url{https://www.w3.org/TR/odrl-model/}}, which was released in February 2018, is a general rights language, which can be used to define rights to or to limit access to digital resources.     
In principle any of these languages could be used to encode usage policies, after the necessary auxiliary ontologies have been integrated. In SPECIAL we developed our usage policy language using OWL2, and select language constructs carefully in order to achieve an optimal trade-off between expressiveness and computational complexity. Our decision was influenced by the fact that we could benefit from out of the box reasoning offered by OWL reasoners (e.g. HermiT and FaCT++).

As for transparency with respect to data processing, relevant work primarily relates to the re-purposing of existing logging mechanisms as the basis for personal data processing transparency and compliance \cite{bonatti2017transparent}. Many of the works use a secret key signing scheme based on Message Authentication Codes (MACs) together with a hashing algorithm to generate chains of log records that are in turn used to ensure log confidentiality and integrity \cite{bellare1997forward} (cf., \cite{bonatti2017transparent} for a summary of existing approaches). MACs are themselves symmetric keys that are generated and verified using collision-resistant secure cryptographic hash functions. However, only a few works \cite{pulls2013distributed,sackmann2006personalization} focus on personal data processing.  Sackmann et al \cite{sackmann2006personalization} demonstrate how a secure logging system can be used for privacy-aware logging. Additionally, they introduce the ``privacy evidence'' concept and discuss how such a log could be used to compare data processing to the user's privacy policy. \citet{samavi2018publishing} propose an ontology that can be used to model personal data processing and demonstrate how SPARQL with limited RDFS reasoning can be used for query-based privacy auditing. 
A distributed architecture to manage access to personal data based on blockchain technology has been proposed by \citet{zyskind2015decentralizing}. The authors discuss how the blockchain data model and Application Programming Interfaces (APIs) can be extended to keep track of both data and access transactions. More recently, \citet{sutton2017blockchain} propose an extension of blockchain technology with \emph{Linked Data} to create tamper-proof audit logs and non-repudiation. Nonetheless, very little research has been conducted into the suitability of such blockchain-based solutions in an industry context. 

Additionally, just focusing on the representation, there exists a number of general event vocabularies such as the \emph{Event}\footnote{Events, \url{http://motools.sourceforge.net/event/event.html}} ontology and the \emph{LODE}\footnote{LODE, \url{http://linkedevents.org/ontology/}} ontology \cite{rinne2013event} that could potentially be used to model privacy-aware data processing \emph{events}. However, these ontologies do not consider the particularities and requirements of the data processing and sharing events considered herein. 
The management of events for business process compliance monitoring and process mining \cite{ly2015compliance}  
can be seen as orthogonal work.

As for GDPR compliance, recently the Information Commissioner's Office (ICO) in the UK \cite{ICO2017}, Microsoft \cite{MicrosoftTrustCenter2017}, and Nymity \cite{Nymity} have developed compliance tools that enable companies to assess the compliance of their applications and business processes by completing a predefined questionnaire. In addition there has been a body of work looking at modelling GDPR concepts and obligations \cite{pandit2018gdprtext,de2019odrl}, in a manner that enables compliance checking beyond consent and transparency. 

In this paper, we propose vocabularies that can be used to record both usage policies and data processing and sharing events in a manner that supports automatic compliance checking. One of the primary differentiators being that both our policy language and our event log have been derived from the legal inquiry process used to assess if personal data processing and sharing complies with the GDPR.

\section{Conclusions}
\label{s:conclusions}

In this paper, we introduced the initial suite of ontologies and vocabularies, developed within the SPECIAL project, that can be used by companies to record consent and data processing and sharing events in the manner that enables the compliance of existing Line of Business and Business Intelligence applications to be checked automatically. 
In addition to providing an overview of the resources and how they fit into the SPECIAL consent, transparency and compliance architecture, we also described how our SPECIAL-K Apache Kafka based big data platform can be used for automatic compliance checking.
We proposed a synthetic SPECIAL compliance benchmark, which is designed on the basis of well-identified choke points (challenges) that could affect the performance of SPECIAL-K and similar systems. 
Finally, we used the SPECIAL compliance benchmark to evaluate the SPECIAL-K compliance checking, using synthesised data.

Future work includes refining the vocabularies based on additional use cases and demonstrating their effectiveness in various business settings. In addition, we plan to continue optimising our SPECIAL compliance checking platform.

\section*{Acknowledgments} 
This work was supported by the European Union's Horizon 2020 research and innovation programme under grant 731601. The authors are grateful to all of SPECIAL’s partners, without their contribution this project and its results would not have been possible.

\bibliographystyle{abbrvnat}
\bibliography{main}

\end{document}